\documentclass[11pt,prd,preprintnumbers,eqsecnum,superscriptaddress,nofootinbib]{revtex4}
\usepackage[usenames,dvipsnames]{xcolor}
\usepackage{graphics,epsfig}
\usepackage{graphicx}
\usepackage{color}
\usepackage{amssymb,amsmath,amsfonts}
 \usepackage{amsmath}
\usepackage{amssymb}
\usepackage{hyperref}
\usepackage{epstopdf}

\def\td#1{\tilde{#1}}
\def\check{ \maltese {\bf Check!}}

\begin{document}
\def\td#1{\tilde{#1}}
\def\check{ \maltese {\bf Check!}}

    \hfill\parbox{5cm} { }

    \vspace{25mm}

\begin{center}
{\Large \bf Vector Meson Gravitational Form Factors and Generalized Parton Distributions at finite temperature within the soft-wall AdS/QCD Model }

\vskip 1. cm
  {Minaya Allahverdiyeva $^{a}$\footnote{minaallahverdiyeva@ymail.com}} and
  {Shahin Mamedov $^{a,b}$\footnote{corresponding author: sh.mamedov62@gmail.com  }},

\vskip 0.5cm

{\it $^a\,$ Institute of Physics, Azerbaijan National Academy of Sciences,

    H. Cavid avenue. 33, Baku, AZ-1143}
\\ \it \indent $^b$ Institute for Physical Problems, Baku State University,

 Z. Khalilov street 23, Baku, AZ-1148, Azerbaijan
Azerbaijan\\
\end{center}

\thispagestyle{empty}

    \vskip2cm

\centerline{\bf Abstract} \vskip 4mm
We investigate the vector meson's gravitational form factors (GFFs) and generalized parton distributions (GPDs) at finite temperatures within the soft-wall AdS/ QCD model. The plotted soft-wall GFFs graphs at zero temperature are close to the hard-wall results \cite{Abidin:2008ku}. The dependence on temperature of the GFFs and GPDs is studied using the thermal dilaton approach in the soft-wall model. Plots of thermal  GFFs show that they decrease in temperature increase. Also, the gravitational radius of $\rho$ meson decreases in temperature growth and becomes zero around the critical temperature $T_c=0.2  {\rm\ GeV}$. GPDs at zero skewness are plotted and analyzed at zero and finite temperature cases.

 \vspace{2cm}

\section{Introduction}

The GFFs and GPDs are among the main quantities giving information about the spatial distribution of partons inside hadrons. In recent years, GFFs have attracted great interest in the hadron physics community and different models are applied for the calculation of these functions. The GFFs are defined as the form factors of the energy-momentum tensor (EMT), which describe the elastic interaction between the graviton and the particles \cite{Abidin:2008ku}. Knowing the EMT enables us to find GPDs describing the distribution of partons inside the hadron and investigate the hadron-graviton vertex. GPDs can be interpreted as amplitudes for removing a parton from a hadron and replacing it with one of different momentum and GPDs are an essential metric of hadron structure. A connection between the moments of the GPDs with GFFs is established. In particular, it was shown the total angular momentum carried by partons is measured by one of the GFFs~\cite{Ji:1996ek,Diehl:2003ny}. Simultaneously, during the last decades, a newly developed theory of gauge-gravity duality (or AdS/CFT correspondence) has been applied in solving particle physics problems. The QCD models based on this duality are called the AdS/QCD models. Two main models of AdS/QCD, hard- and soft-wall models, were applied for calculating the GFFs and GPDs of hadrons within the AdS/QCD models and turned out effective ones for this study. Authors in Ref.~\cite{Abidin:2008ku, Abidin:2008} have considered the GFFs for the vector and axial vector mesons within the hard-wall model of AdS/QCD. These form factors for the nucleons were studied within the both hard- and soft-wall models in Ref.~\cite{Abidin:2009}. The soft- and hard-wall GFFs results for the nucleons were obtained in good agreement with the Kelly empirical fit. The GFFs for the pions within the AdS/QCD framework were considered in Ref.~\cite{Abidin:2008}. In Ref.~\cite{AbidinCarlson} the GFFs of nucleons and $\rho$ mesons were applied for the study of the momentum densities inside a hadron. In further studies, a relation between the GPDs of the hadrons and GFFs was found within the AdS/QCD models. GPDs in the soft-wall model were considered a matching procedure of sum rules relating the electromagnetic form factors (EFFs) to GPDs in the zero skewness case in Ref.~\cite{VegaGPD} and the non-zero skewness case in Refs.~\cite{VegaGPDs, VegaGPDss}. A similar analysis was done within the hard-wall model in Ref.~\cite{Vega}. In Ref. \cite{Sharma} author investigates GPDs Mellin moments and transverse impact parameter GPDs within the soft-wall model. In Ref.~\cite{Mondal2} the longitudinal momentum densities for nucleons were analyzed by use of a relation between GPDs and electromagnetic form factors calculated in the AdS/QCD framework. In Ref.~\cite{Mondal} the transverse densities for deuteron and in Ref.~\cite{Navdeep} the GPDs of pion were investigated by use of AdS/QCD. In Refs.~\cite{WieXie1, WieXie2, WieXie3} the holographic expressions of GFFs were applied for the study of the elastic proton-proton, pion-proton, and pion-pion scattering processes. GPDs and GFFs we studied in Ref.~\cite{Teramond} and in Ref.~\cite{Fujita} within the light-front model and top-down approach of holographic QCD. In Ref.~\cite{Traini} nucleon GPDs were studied within the soft-wall model with the aim to extend this approach from the forward to the off-forward region. Hadron and deuterium structure functions in the framework of the hard-wall model were studied in \cite{Agozzino}. Flavour form factors, charge and magnetization densities of the nucleons within the hard-wall model were investigated in \cite{Mondalh}.

Recently, in Ref.~\cite{Lyubovitskij:20192, Lyubovitskij:2019} the soft-wall model was extended to the finite temperature case by making the dilaton field a thermal one and the profile functions of free mesons and baryons were found in a such thermal bath. This model has opened new opportunities in the study of the hadron’s phenomenological quantities such as coupling and decay constants, form factors, etc. at finite temperatures \cite{Lyubovitskij:2020, ShahinM1, ShahinM2}. One of these opportunities is to investigate the GFFs and GPDs at finite temperature cases and plot the temperature dependencies of these quantities. This could give us information about the change in the spatial distribution of the partons under the influence of the thermal bath. The $\rho$ meson GFFs and GPDs investigation are still under interest (\cite{Aliev, Alharazin, Sun, Sun2, Epelbaum, Selyugin}) and we aim to extend it to the finite temperature case using known holographic results for these functions at zero temperature case and the thermal dilaton soft-wall AdS/QCD model.
 
This work is organized in the following sections. In Sec.~\ref{sec:model} we introduce the soft-wall model. In Sec.~\ref{vectorpf} we present the solution to the equation of motion for the graviton and vector meson. In Sec.~\ref{sec:ff}  we consider the GFFs at finite temperature. In Sec.~\ref{GPD} we calculate the GPD at a finite temperature. In Sec.~\ref{numanal} we present our numerical results. InSec.~\ref{summary} we summarize our results.

%
\section{The soft-wall model}           \label{sec:model}
%

The gravity background for the model is the AdS space and the metric in Poincare coordinates has a form:
\begin{equation}
ds^2=g_{MN} dx^M dx^N=\frac{L^2}{z^2}\,\bigg(g_{\mu\nu}dx^{\mu}dx^{\nu}-dz^2\bigg) \label{2.1},
\end{equation}
where  $\eta_{\mu\nu}=diag(1,-1,-1,-1)$, $\mu,\nu=0,1,2,3$ is the Minkowski metric, $z$ is the holographic coordinate varying from $0$ to $\infty$ and $L$ is the AdS curvature radius.

Vector field $V$, which describes the vector mesons within the AdS/QCD models, is introduced by the sum of two gauge fields $A_{L, R}$ corresponding to the $SU(2)_L$ and  $SU(2)_R$  subsymmetries of the  $SU(2)_L \times SU(2)_R$ symmetry of the model. $V$ is composed as $V= \left(A_{L}+A_{R}\right)/2$ and interacts with the perturbation of the metric. The interaction vertex is described by the vector mesons GFFs. In addition to the gauge fields, we have the dilaton field $\varphi(z)=\kappa^2 z^2$, which does not describe any particle states and is essential for the Regge behaviour of the vector meson spectrum. In the soft-wall model case, the field is also responsible for the breaking of the model's chiral and conformal symmetries. 

The 5D action for the model is the sum of the gravity and gauge field parts:
 \begin{equation}
S=\int d^{5}x\,\sqrt{g}\,e^{-\varphi(z)}\{R+12-\frac{1}{4g_{5}^2}\,{\mathrm{Tr}}(F_{L}^2+F_{R}^2)\},\label{2.2}
\end{equation}
where $g=|det g_{MN}|$, $F_{L,R}^{MN}=\partial^MA_{L,R}^N-\partial^NA_{L,R}^M-i[A_{L,R}^M,A_{L,R}^N]$, $A_L^M=A_L^{a,M}t_L^a$, $A_R^M=A_R^{a,M}t_R^a$. ${\mathrm{Tr}}(t^a_L t^a_R)=\delta^{ab}/2$ and $t_L^a$ and $t_R^a $ are the generators of the ${SU}(2)_L$ and ${SU}(2)_R$ chiral groups respectively, $R$ is the Ricci scaliar.
The finiteness of this action is guaranteed by the $e^{-\varphi(z)}$ factor.

Let us consider the soft-wall model with thermal dilaton, which was built in Refs.~\cite{Lyubovitskij:20192, Lyubovitskij:2019}, and present some necessary formulas of this model. Thermalization of the model is done by introducing the thermal dilaton field $\varphi_T(z) = K_T^2 z^2$ and this turned out equivalent to introducing the AdS-Schwarzschild background at low temperature:
\begin{equation}
ds^2 = e^{2 A(z)} \, 
\biggl[ f_T(z) dt^2 - (d\vec{x})^2 - \frac{dz^2}{f_T(z)} \biggr] \label{5.1},
\end{equation}
where $f(z) = 1 - z^4/z_h^4$, $A(z) = \log(R/z)$, $z_h$ is the position of the black hole horizon and $x=(t,\vec{x}\,)$ is the set of Minkowski coordinates. The temperature of the 4D theory is defined as Hawking temperature $T = 1/(\pi z_h)$ of the black hole.

Tortoise coordinate $r$ in the model is related with the $z$ coordinate and the metric (\ref{5.1}) in terms of $r$ coordinate gets a form:
\begin{equation}
ds^2= e^{2A(r)}f^{3/5}(r)\left(dt^2 - \frac{d\overrightarrow{x}^2}{f(r)} -dr^2 \right). \label{5.3}
\end{equation}
Temperature dependence of the thermal dilaton field $\varphi(z,T)$ of the soft-wall model was established in Ref. \cite{Lyubovitskij:20192} in the form:
\begin{equation}
\varphi_T(r) = K_T^2 r^2, \label{5.4}
\end{equation}
where
\begin{equation}
K_T=\sqrt{(1 + \rho_T) \kappa }, \label{5.5}
\quad 
\rho_T = \biggl(\frac{9 \alpha \pi^2}{16}
\,+\, \delta_{T_1}\biggr) \frac{T^2}{12 F^2}
\,+\, \delta_{T_2} \biggl(\frac{T^2}{12 F^2}\biggr)^2
\,+\,{\cal O}(T^6).
\end{equation}
Here $F = \kappa \frac{\sqrt{3}}{8} \,$ is the pion a decay constant and $\delta_{T_1}$ and $\delta_{T_2}$ are the constants defined by the number of quark flavors $N_f$:
\begin{equation}
\delta_{T_1}  =  - \frac{N_f^2-1}{N_f}     \,, \quad  \label{5.6}
\delta_{T_2} \,=\, - \frac{N_f^2-1}{2 N_f^2}. 
\end{equation}

\section{graviton bulk-to-boundary propagator and vector meson profile function} \label{vectorpf}
\subsection{The Graviton bulk-to-boundary propagator}

The 4D metric $g_{\mu\nu}$ in (\ref{2.1}) is linearly perturb around its static solution $g_{\mu\nu}=\frac{L^2}{z^2} \eta_{\mu\nu}$: 
\begin{equation}
g_{\mu\nu}=\frac{L^2}{z^2}\left(\eta_{\mu\nu}+h_{\mu\nu}(x,z)\right).\label{3.2}
\end{equation}
The gravity part of the action in the second order of the $h_{\mu\nu}$ perturbation becomes \cite{Abidin:2009}:
\begin{eqnarray}
S_{GR}=-\int d^5 x\, \frac{e^{-2\kappa^2z^2}}{4z^3}  \left(h_{\mu\nu,z}{h^{\mu\nu}}_{,z}+h_{\mu\nu}\Box h^{\mu\nu} \right),    \label{3.3}
\end{eqnarray}
where the transverse-traceless gauge conditions $\partial^\mu h_{\mu\nu}=0$ and $h^\mu_\mu=0$ were imposed. The perturbation $h_{\mu\nu}$ in momentum space will be written in the $h^{\mu\nu}(q,z)=h^{\mu\nu}(q)h(q,z)$ form. The linearized Einstein equations obtained from the action Eq.~(\ref{3.3}) give the following equation for the $h(q,z)$ part:
\begin{eqnarray}
\left[ \partial_z\left(\frac{e^{-2\kappa^2z^2}}{z^3}\partial_z\right) +\frac{e^{-2\kappa^2z^2}}{z^3}q^2\right] h(q,z)=0.    \label{3.4}
\end{eqnarray}
The non-normalizable solution $h(q,z)$ is called graviton bulk-to-boundary propagator and at $q^2=-Q^2$ domain it is expressed by the Tricomi hypergeometric function $U$ \cite{Brian}:
\begin{eqnarray}\ \label{3.5}
H(Q,z)=\Gamma(a+2) U(a,-1;2\kappa^2z^2)=
\nonumber\\
=&a(a+1) \int_0^1 dx \, x^{a-1} (1-x) \exp\left(\frac{-2\kappa^2z^2 x}{1-x}\right),    
\end{eqnarray}
where $H(Q,z)\equiv h(q^2=-Q^2,z)$ and $a=Q^2/8k^2$. This solution was obtained by imposing boundary conditions $H(q,\epsilon)=1$, $(\epsilon\rightarrow0)$ and  $\partial_{z}H(p,z))|_{z=\infty}=0$.

The action for the graviton at finite temperature in the second-order perturbation in the thermal dilaton approach will be obtained from the zero temperature action (\ref{3.3}) by replacing $\kappa\rightarrow K_T$, $z\rightarrow r$ and the graviton equation of motion has the form as in Eq.~(\ref{3.4}):
\begin{eqnarray}\ \label{3.6}
\left[ \partial_r\left(\frac{e^{-2K_T^2r^2}}{r^3}\partial_r\right) +\frac{e^{-2K_T^2r^2}}{r^3}q^2\right] h(q,T,r)=0,    
\end{eqnarray}
Accordingly, the graviton bulk-to-boundary propagator at finite temperature, which is the non-normalizable solution of (\ref{3.6}), gets a form of the solution at zero temperature (\ref{3.5}) with the replacement $a\rightarrow a_T$:
\begin{eqnarray}\ \label{k}
H(Q,T,r)=\Gamma(a_{T}+2) U(a_{T},-1;2K_T^2r^2)= 
\nonumber \\
=a_{T}(a_{T}+1) \int_0^1 dx \, x^{a_{T}-1} (1-x) \exp\left(\frac{-2K_T^2r^2 x}{1-x}\right),
 \end{eqnarray}
where $a_T=\frac{Q^2}{8K_T^2}$. Another way to obtain graviton bulk-to-boundary propagator is to take mass $m=0$ and spin $J=2$ in the general expression for the profile function for the spin $J$ fields at finite temperature found in Ref \cite{Lyubovitskij:20192}. The result is the same with (\ref{k}).

\subsection{Vector meson profile function} 

The equation of motion for the $\psi_n(z)$ mode in the Kaluza-Klein decomposition $V_\mu(q,z)=\Sigma V_\mu^n (q)\psi_n(z)$ of the vector field is obtained from the action Eq.~(\ref{2.2}) and accepts the form \cite{Karch:2006pv}:
\begin{eqnarray}
\partial_z\left[\frac{1}{z}e^{-\kappa^2z^2} \partial_z\psi_n(z) \right] + m^2_\rho\,
\frac{1}{z}\, e^{-\kappa^2z^2}\psi_n(z) = 0 \ .  \label{3.8}
\end{eqnarray}
By the substitution
 \begin{eqnarray}\label{3.7}
\psi_n(z) = e^{\kappa^2z^2}\sqrt{z}\, \Psi_n(z)
\end{eqnarray}
this equation has been brought into the form of  Schr\"{o}dinger  equation ~\cite{Karch:2006pv} and the solution for the first Kaluza-Klein mode, i.e. $\rho$ meson's profile function $\psi_n(z)$, is expressed in terms of the Laguerre polynomials  $L^1_n$:
\begin{eqnarray}
\psi_n(z) = \kappa^2z^2\sqrt{\frac{2n!}{(n+1)!}} \, L^1_{n}(\kappa^2 z^2) \ .  \label{3.9}
\end{eqnarray}
The mass spectrum corresponding to these states is $m^2_n=4k^2(n+1), n=0,1,2,3,...$.
The profile function for the vector meson  at finite temperature was found in the form ~\cite{Lyubovitskij:20192}:
\begin{equation}
\psi(r,T) = \sqrt{\frac{2 \Gamma(n+1)}{\Gamma(n+m+1)}} \ \label{3.10}
K_T^{m+1}
\ r^{m+1/2} \ e^{- \varphi_T(r)/2} \ L_n^m(\kappa^2 r^2) \,.
\end{equation}

\section{Gravitational Form Factors of Vector Meson}    \label{sec:ff}

Energy-momentum stress tensor $T^{\mu\nu}$ is defined as a variation of the actions over the metric tensor $g_{\mu\nu}$, which in the weak perturbation (\ref{3.2}) case is reduced to the variation over the $h_{\mu\nu}$ perturbation:
\begin{equation}
    T^{\mu\nu}=-\frac{2}{\sqrt{-g}} \frac{\delta S}{\delta{h_{\mu\nu}}}.
    \label{4.1}
\end{equation}
 In the Lagrangian formulation of general relativity, the source for the   $T^{\mu\nu}$ tensor is the metric tensor $g^{\mu\nu}$. The energy-momentum tensor is conserved $\left(q_{\mu}T^{\mu\nu}=0\right)$. Due to the $\mu	\Leftrightarrow\nu$ symmetry of $T_{\mu\nu}$ and its conservation, the number of independent components of $T_{\mu\nu}$ reduces to six.  This tensor can be decomposed into the transverse-traceless part  $\hat{T}^{\mu\nu}$, which has five independent components, and the transverse-not-traceless part $\tilde {T}$, which has one component:
\begin{equation}
    T^{\mu\nu}=\hat{T}^{\mu\nu}+\tilde{T}^{\mu\nu}.
    \label{4.2}
\end{equation}
Transverse-traceless $h_{\mu\nu}$ couples with  $\hat{T}^{\mu\nu}$.

Stress tensor matrix elements of $\rho$ meson $\big<\rho_n^a(p_1)|T^{\mu\nu}(q)|\rho_n^b(p_2)\big>$ can be extracted from the 3-point function
$\big< 0 \big|T \big( {J_a}^\alpha(x)  T^{\mu\nu}(y)  {J_b}^\beta(w)  \big)\big|0\big> $. At the linear perturbation approximation, only the $hVV$ terms in the actions contribute to the 3-point functions
\begin{equation}
\big< 0 \big| {\mathcal T} {J}^\alpha(x) \hat{T}^{\mu\nu}(y){J}^\beta(w)\big| 0 \big>	=
	\frac{  -  2 \, \delta^3 S}{\delta V^0_\alpha (x) \delta h^0_{\mu\nu}(y) \delta V^0_\beta(w)} ,
		\label{4.7}
\end{equation}
where the functional derivative is evaluated at $h^0=V^0=0$.

The transverse-traceless projector was applied in the action to get the conserved and traceless $T^{\mu\nu}$. In the result of variational derivatives from the action the matrix element $\left<\rho_n^a(p_2,\lambda_2)\big|\hat{T}^{\mu\nu}(q)\big|\rho_n^b(p_1,\lambda_1)\right>$ was written in terms of the form factors $A(q^2), \hat C(q^2), D(q^2), \hat F(q^2)$
\begin{eqnarray}
\left<\rho_n^a(p_2,\lambda_2)\big|\hat{T}^{\mu\nu}(q)\big|\rho_n^b(p_1,\lambda_1)\right>=\nonumber \\
\quad (2\pi)^4 \delta^{(4)}(q+p_1-p_2) \, \delta^{ab} \,
	\varepsilon^*_{2 \alpha} \varepsilon_{1 \beta} \nonumber\\[1ex]
\times\bigg[ - A(q^2)\bigg(4 q^{[\alpha} \eta^{\beta](\mu} p^{\nu)}
	+2\eta^{\alpha\beta}p^\mu p^\nu\bigg)				\nonumber\\
\quad-\ \frac{1}{2} \hat C(q^2)
\eta^{\alpha\beta}\bigg(q^2 \eta^{\mu\nu} - q^\mu q^\nu  \bigg) \nonumber\\
\quad +\  D(q^2) \bigg( q^2 \eta^{\alpha(\mu}\eta^{\nu)\beta}
	- 2q^{(\mu}\eta^{\nu)(\alpha}q^{\beta)}
		+ \eta^{\mu\nu} q^\alpha q^\beta \bigg) \nonumber\\
\quad-\ \hat F(q^2) \frac{q^\alpha q^\beta}{m_n^2}
	\bigg( q^2 \eta^{\mu\nu}- q^\mu q^\nu \bigg)\bigg] \,, 
				\label {}
\end{eqnarray}
which are expressed by the invariant functions $Z_1$ and $Z_2$:
\begin{eqnarray}
A(q^2)&=&Z_2   \,,  \nonumber \\
\hat C(q^2)&=&  \frac{1}{q^2}
    \bigg(\frac{4}{3}Z_1+\big( q^2 - \frac{8m_\rho^2}{3}\big)Z_2\bigg), \nonumber \\
D(q^2)&=& \frac{2}{q^2} Z_1+\left( 1- \frac{2m_\rho^2}{q^2} \right)Z_2, \nonumber \\
\hat F(q^2)&=& \frac{4m_\rho^2}{3q^4}  \left( Z_1 - m_\rho^2Z_2 \right),     \label{4.12}
\end{eqnarray}
The  $Z_1$ and $Z_2$ invariant functions in the soft-wall model, have a similar form, as within the  hard-wall model~\cite{AbidinCarlson}:
\begin{eqnarray}
Z_1(Q)&=&\int \frac{dz}{z} e^{-\kappa^2z^2} H(Q,z)\partial_z\psi_n(z) \partial_z\psi_n(z) \,,
            \nonumber \\
Z_2(Q)&=&\int \frac{dz}{z} e^{-\kappa^2z^2} H(Q,z) \psi_n(z) \psi_n(z) \,.    \label{4.13}
\end{eqnarray}

The gravitational RMS (root mean square) radius is defined through taking the derivative from the gravitational form factor $A(q^2)$:
\begin{equation}
\left\langle r^2 \right\rangle_{\rm grav} = -6 \left. \frac{\partial A}{\partial Q^2} \right|_{Q^2=0} . \label{4.14}
\end{equation}
Following the hard-wall model ~\cite{Abidin:2008ku}, in order to calculate $\left\langle r^2 \right\rangle_{\rm grav}$ radius we expand $H(Q,z)$ at small $Q^2$:
\begin{equation}
H(Q,z)\approx 1+\frac{1}{8k^2}Q^2(1-\gamma +U(0,-1,2k^2z^2)) \label{4.15}.
\end{equation}
Using this expansion the value of the gravitational radius can be calculated, and this gives us the following value of the radius: 
\begin{equation}
\left\langle r^2 \right\rangle_{\rm grav} =-0.27 {\rm\ fm}^2 , \label{4.16}
\end{equation}
which is close to the hard-wall value $0.21 {\rm\ fm}^2$.

The $Z_1$ and $Z_2$ functions at finite temperature will be written in terms of profile $\psi(r, T)$ and $H(Q, T,r)$:
\begin{eqnarray}
Z_1(Q,T)&=&\int \frac{dr}{r}  H(Q,T,r)\partial_r\psi(r,T) \partial_r\psi(r,T) \,,
            \nonumber \\
Z_2(Q,T)&=&\int \frac{dr}{r}  H(Q,T,r)\psi(r,T) \psi(r,T) \,.    \label{5.10}
\end{eqnarray}

Temperature dependence of the gravitational radius can be easily calculated by use of $A(Q, T)=Z_2(Q, T)$:
\begin{equation}
\left\langle r^2(T) \right\rangle_{\rm grav}= -6 \left. \frac{\partial Z_2(Q,T)}{\partial Q^2} \right|_{Q^2=0} . \label{6.11}
\end{equation}

\section{Generalized parton distributions for the $\rho$ meson} \label{GPD}

Since we have the holographic expressions for GFFs, we can calculate the GPDs for the $\rho$ meson using the relation between these functions. For the vector mesons there are five GPDs defined in Ref.~\cite{Abidin:2008ku, Berger, Diehl:2003ny} and denoted by $H_i$: 
 \begin{eqnarray}
&& \int \frac{p^+ dy^-}{2\pi} e^{ixp^+y^-} \times 
						\nonumber \\
&& \quad	\left\langle p_2,\lambda_2 \right|  \bar q(-\frac{y}{2}) \gamma^+  
		q(\frac{y}{2})\left| p_1, \lambda_1 \right\rangle_{y^+=0, y_\perp = 0}
						\nonumber \\
&& = - 2 (\varepsilon_2^* \cdot \varepsilon_1) p^+ H_1(x,\xi,t)
	- \left( \varepsilon_1^+ \, \varepsilon_2^* \cdot q 
		- {\varepsilon_2^+}^* \, \varepsilon_1 \cdot q \right) H_2(x,\xi,t)
						\nonumber \\
&& \quad + \ q\cdot \varepsilon_1 \, q\cdot \varepsilon_2^* \frac{p^+}{m_n^2} H_3(x,\xi,t)
	- \left( \varepsilon_1^+ \, \varepsilon_2^* \cdot q 
		+ {\varepsilon_2^+}^* \, \varepsilon_1 \cdot q \right) H_4(x,\xi,t)
						\nonumber \\
&& \quad +\ \left( \frac{m_n^2}{(p^+)^2} \varepsilon_1^+ \, {\varepsilon_2^+}^*
		+ \frac{1}{3} (\varepsilon_2^* \cdot \varepsilon_1) \right) 2 p^+ \, H_5(x,\xi,t) \ , \label{6.1}
\end{eqnarray}

where $p^+$ is the light-cone momentum $p^+=\frac{p^0+p^3}{\sqrt{2}}$  of the meson, momentum fraction $x=\frac{k^+}{p^+}$ varying $-1\leq x\leq 1$ interval, $k^+$ is the parton’s (quark’s) light-cone momentum. Each of the GPDs has arguments, where $q^+=-2\xi p^+$, $t=q^2$ and $\xi$ is skewness. Integration of $H_i$ on $x$ gives electromagnetic form factors of the meson. Besides this definition, the integrals over the $x H_i(x,\xi ,t)$ are related to GFFs \cite{Abidin:2008ku}:
\begin{align}
\int_{-1}^1 x dx \, H_1(x,\xi,t) &=A(t) - \xi^2 C(t) + \frac{t}{6m_\rho^2} D(t)	,	\nonumber \\
\int_{-1}^1 x dx \, H_2	(x,\xi,t) &= 2 \left( A(t)+B(t) \right)		,	\nonumber \\
\int_{-1}^1 x dx \, H_3(x,\xi,t) &= E(t) + 4 \xi^2 F(t)				,	\nonumber \\
\int_{-1}^1 x dx \, H_4(x,\xi,t) &= -2 \xi D(t)					,	\nonumber \\
\int_{-1}^1 x dx \, H_5(x,\xi,t) &=  + \frac{t}{2m_\rho^2} D(t)			\label{6.2}	.	
\end{align}

Four of the right-hand side functions were given in (\ref{4.12}), and the other two were found in Ref.~\cite{Abidin:2008ku} vanishing ones:
\begin{equation}
B(t)=E(t)=0. \label{6.3}
\end{equation}

    The (\ref{6.2}) relations allow us to write explicit expressions of $H_i (x,\xi,t)$. To this end, the integration domain in (\ref{6.2}) can be reduced to $0\leq x\leq1$ \cite{VegaGPD, Navdeep, Berger}:
    \begin{equation}
\int_{-1}^1dxH_i(x,\xi,t)=\int_{0}^1dx H_i(x,\xi,t)+\int_{0}^1dx H_i(-x,\xi,t)=\int_{0}^1 dx H_i^v(x,\xi,t) . \label{6.5}
\end{equation}
Now let us write the GFFs in (\ref{4.13})  using the integral representation of $H(Q, z)$ in (\ref{3.5}). To this end let us reduce the (3.7) integral representation of $H(Q, r)$ to the following form:   
   \begin{eqnarray}
H(Q,z)=4\int_{0}^1dx  \kappa^4 z^4\frac{x^{a+1}}{(1-x)^3}e^{\frac{-2\kappa^2z^2x}{1-x}}. 
\end{eqnarray}
In terms of this representation  of  $H(Q, r)$  the GFFs will accept the form:    
  \begin{eqnarray}
Z_1(Q)=32\int_{0}^1dx\int_{0}^\infty dz \kappa^8 z^5\frac{x^{a+1}}{(1-x)^3}e^{\frac{-\kappa^2z^2(1+x)}{1-x}},
\nonumber \\
Z_2(Q)=8\int_{0}^1dx\int_{0}^\infty dz \kappa^8 z^7\frac{x^{a+1}}{(1-x)^3}e^{\frac{-\kappa^2z^2(1+x)}{1-x}}. \label{6.3}
\end{eqnarray}
Writing this representation in (\ref{4.12}) the GPDs for the $\rho$ meson one can obtain by:
\begin{multline}
H_1^v(x,0,t) =\int_{0}^\infty dz \, \frac{4\kappa^8z^5x^{-a}}{3m^2_{\rho}t(1-x)^3}e^{\frac{-\kappa^2z^2(1+x)}{1-x}}(t^2z^2+t(8+4m^2_{\rho}z^2)),             \\
H_2^v(x,0,t) =\int_{0}^\infty dz \, \frac{16\kappa^8z^7x^{-a}}{(1-x)^3}e^{\frac{-\kappa^2z^2(1+x)}{1-x}}, \\
H_5^v(x,0,t) = \int_{0}^\infty dz \, \frac{4\kappa^8z^5x^{-a}}{m_\rho^2(1-x)^3}e^{\frac{-\kappa^2z^2(1+x)}{1-x}}(8-2m^2_{\rho}z^2+tz^2).		\\	
\label{6.6}			
\end{multline}
Similar explicit expressions for GPDs were obtained for pions in Ref. \cite{Navdeep} and for nucleons in Refs. \cite{VegaGPD, VegaGPDss, Mondal2, Sharma}.
At finite temperature case  the integral representation of $Z_1(Q, T)$ and $Z_2(Q, T)$ functions can be written  replacing k by $K_T$ and $z$ by $r$
\begin{eqnarray}
Z_1(Q,T)=32\int_{0}^1dx\int_{0}^\infty dr K_T^8 r^5\frac{x^{a_T+1}}{(1-x)^3}e^{\frac{-K_{T}^2r^2(1+x)}{1-x}},
\nonumber \\
Z_2(Q,T)=8\int_{0}^1dx\int_{0}^\infty dr K_T^8 r^7\frac{x^{a_T+1}}{(1-x)^3}e^{\frac{-K_{T}^2r^2(1+x)}{1-x}}. \label{6.7}
\end{eqnarray}
and the explicit forms of thermal $H_i^v$ GPDs obtain following forms:
\begin{multline}
H_1^v(x,0,t,T) =\int_{0}^\infty dr \, \frac{4K_T^8r^5x^{-a_T}}{3m^2_{\rho}t(1-x)^3}e^{\frac{-K_T^2r^2(1+x)}{1-x}}(t^2r^2+t(8+4m^2_{\rho}r^2)),            \\
H_2^v(x,0,t,T) =\int_{0}^\infty dr \, \frac{16K_T^8r^7x^{-a_T}}{(1-x)^3}e^{\frac{-K_T^2r^2(1+x)}{1-x}},  \\		
H_5^v(x,0,t,T) = \int_{0}^\infty dr \, \frac{4K_T^8r^5x^{-a_T}}{m_\rho^2(1-x)^3}e^{\frac{-K_T^2r^2(1+x)}{1-x}}(8-2m^2_{\rho}r^2+tr^2).		\\	
\label{6.8}			
\end{multline}

\section{NUMERICAL ANALYSIS} \label{numanal}

\begin{figure*}[htbp]
\begin{minipage}[c]{0.98\textwidth}
\includegraphics[width=7.5cm,clip]{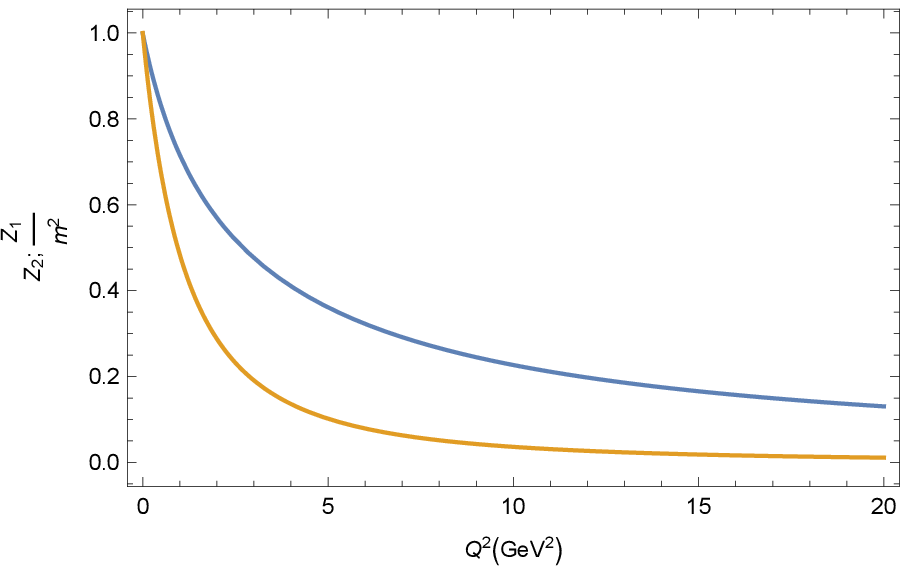}
\end{minipage}
\caption{Plot of $Z_2$ (orange curve) and $Z_1/m^2$ (blue curve), with momentum transfer for the $\rho$ mesons.}
\label{Z1Z_2}
\end{figure*}

\begin{figure*}[htbp]
\begin{minipage}[c]{0.98\textwidth}
{(a)}\includegraphics[width=7.5cm,clip]{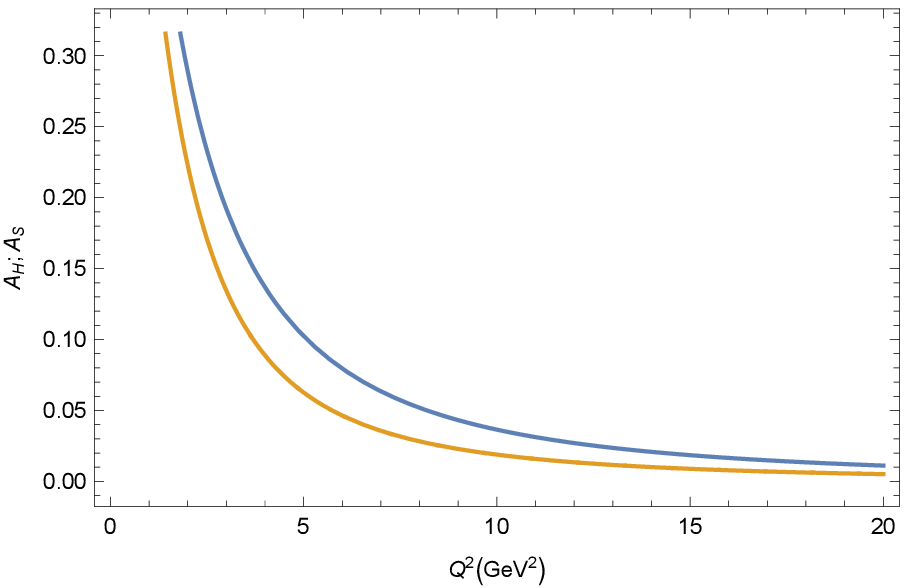}
{(b)}\includegraphics[width=7.5cm,clip]{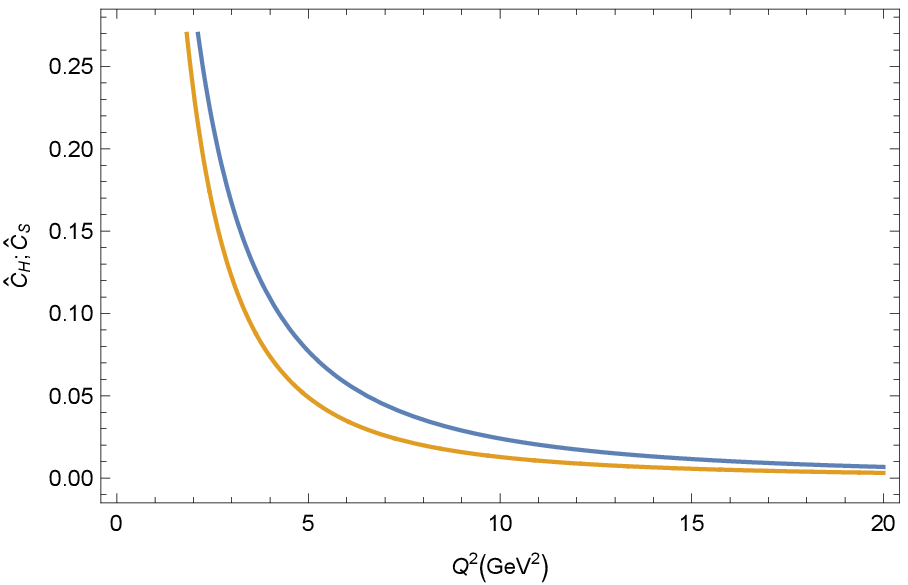}

{(c)}\includegraphics[width=7.5cm,clip]{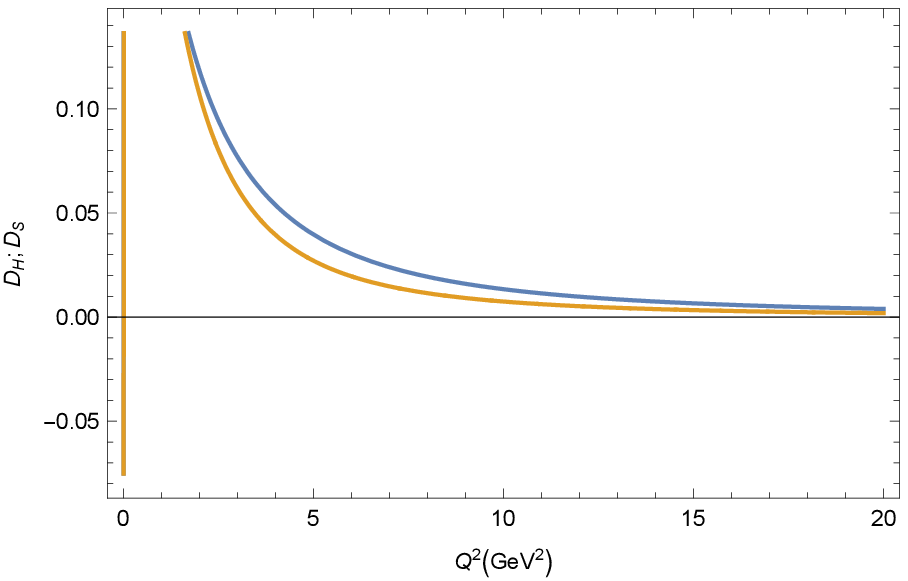}
{(d)}\includegraphics[width=7.5cm,clip]{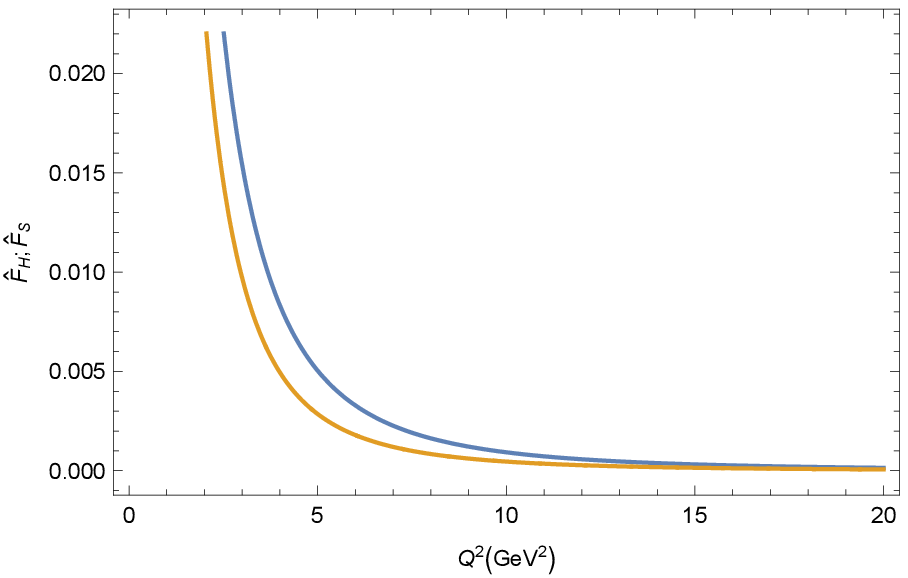}
\end{minipage}
\caption{$A(q^2)$ (a), $\hat C(q^2)$ (b), $D(q^2)$ (c), $\hat F(q^2)$ (d) are the gravitational form factors in the soft-wall model (orange curve) and hard-wall model (blue curve).}
\label{ACD_F}
\end{figure*}

\begin{figure*}[htbp]
\begin{minipage}[c]{0.98\textwidth}
{(a)}\includegraphics[width=7.5cm,clip]{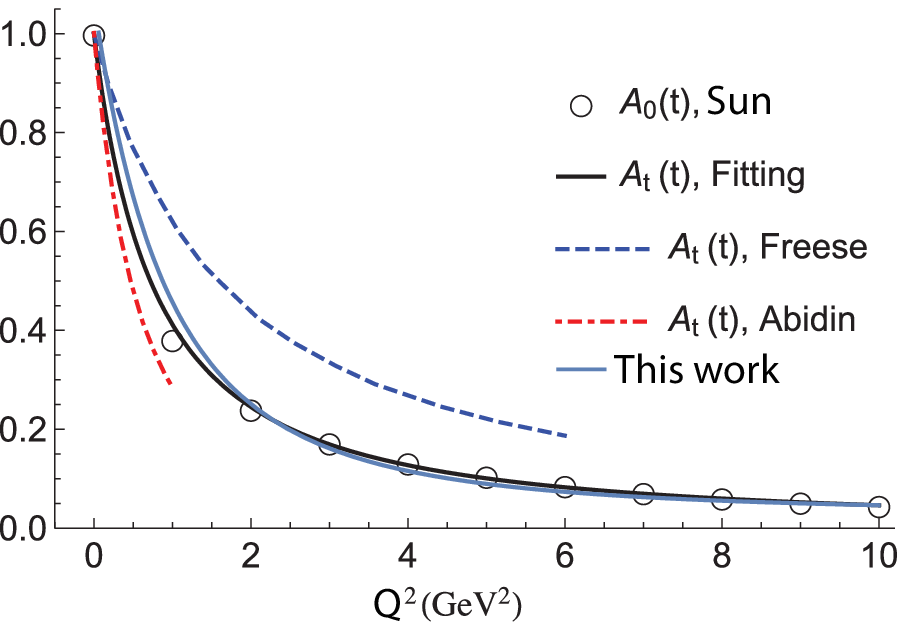}
{(b)}\includegraphics[width=7.5cm,clip]{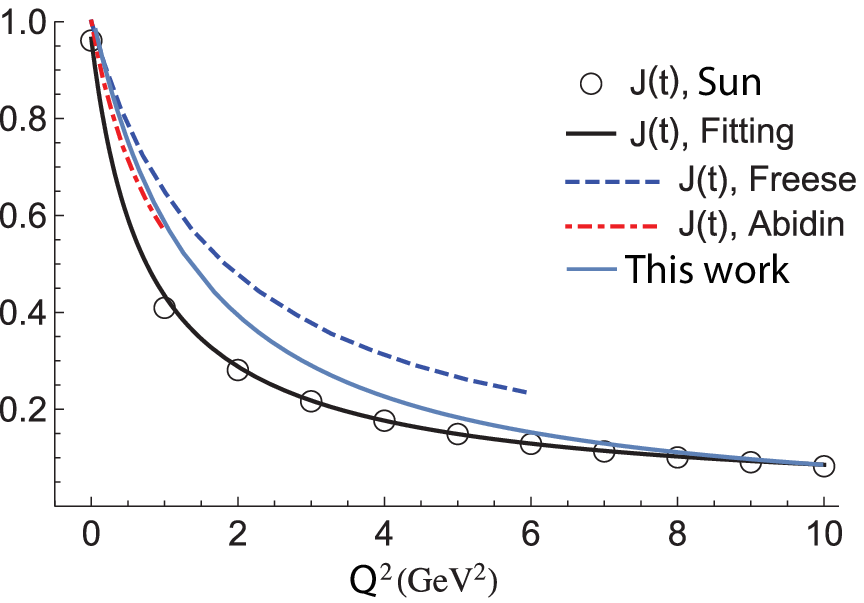}
\end{minipage}
\caption{Plot (a) and (b) are gravitational form factors. The blue lines are the model results. The red dot-dashed lines are results from the AdS/QCD approach by Abidin etc. \cite{Abidin:2008ku}, the empty circles are results from \cite{Sun} and the blue dashed lines are results from the NJL model by Freese etc. \cite{Adam}.  The solid lines are parametric fitting.}
\label{figs}
\end{figure*}

For nucleons, the soft-wall model results at zero temperature were compared with hard-wall ones in \cite{WieXie2}. It is reasonable to make such a check for the $\rho$ meson as well. To this end, in Fig.~\ref{Z1Z_2} we plot graphs for the $Z_2$ and $Z_{1}/m^2$ functions found within the soft-wall model and compare them with the hard-wall model ones plotted in \cite{WieXie2}. The soft- and hard-wall models’ behaviours of these functions are close. Plot and comparison for the $A(q^2), \hat C(q^2), D(q^2), \hat F(q^2)$ form factors graphs obtained within these holographic models have been done in Fig.~\ref{ACD_F} and the closeness of the graphs in both models has been observed. This means the bottom-up approach holographic models are useful ones for the GFFs studies. The shape of the dependence plotted in these graphs is typical for the form factors of hadrons. Moreover, a comparison of the $A(-t=Q^2)$ form factor obtained within the soft-wall model (this work) and hard-wall model \cite{Abidin:2008ku} with the results of the light front constituent model \cite{Sun} and Nambu—Jona-Lazinio (NJL) model \cite{Adam} was given in Fig.~\ref{figs}. We find a good agreement between the results obtained within the both AdS/QCD models and the results of these non-holographic models. The plot in Fig.~ref{figs} were taken from \cite{} and have been added by our soft-wall graphs.

\begin{figure}[htbp]
\includegraphics[width = 0.4\textwidth]{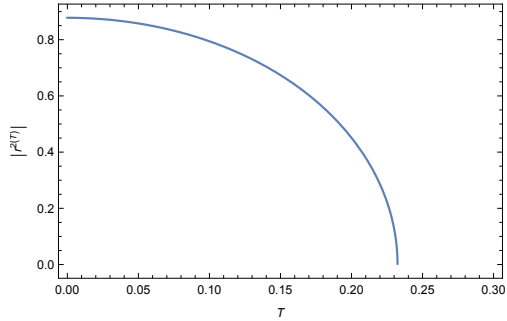}
\caption{Plot of gravitational radius for the $\rho$ mesons.}
\label{R_2}
\end{figure}

 At the finite temperature case, the plot for the $T$ and $Q^2$ dependencies of the $Z_1$ and $Z_2$ functions are given in Fig.~\ref{Soft_Z1Z2} (a) and (b), correspondingly. The $Q^2$ dependence in these plots shows a decrease of these functions in the $0<Q^2<1 {\rm\ GeV}$ region, as was at the zero temperature case. Both functions fall on temperature increasing, though $Z_2$ is less sensitive to the change in the momentum transfer than $Z_1$. These graphs show that $Z_1$ form factor vanishes at the $Q^2 = 0$ value and near the critical temperature $T_c\approx 0.2 {\rm\ GeV}$ for the confinement-deconfinement phase transition. $Z_2$ form factor also vanishes at $T=T_c$ and this vanishing temperature weakly depends on $Q^2$. Such temperature dependency is typical in form factor and coupling constant studies within this soft-wall model \cite{Lyubovitskij:20192, Lyubovitskij:2019, Lyubovitskij:2020} and the vanishment of these quantities at $T=T_c$ was interpreted as a meson splitting into quarks and ending meson state in the result of high temperature \cite{ShahinM1, ShahinM2}. In Fig.~\ref{R_2} a modulus of the RMS gravitational radius dependence on temperature has been plotted. As the figure shows, this radius also decreases on temperature increases and becomes zero around the $T_c$ temperature. 
 
\begin{figure*}[htbp]
\begin{minipage}[c]{0.98\textwidth}
{(a)}\includegraphics[width=7.5cm,clip]{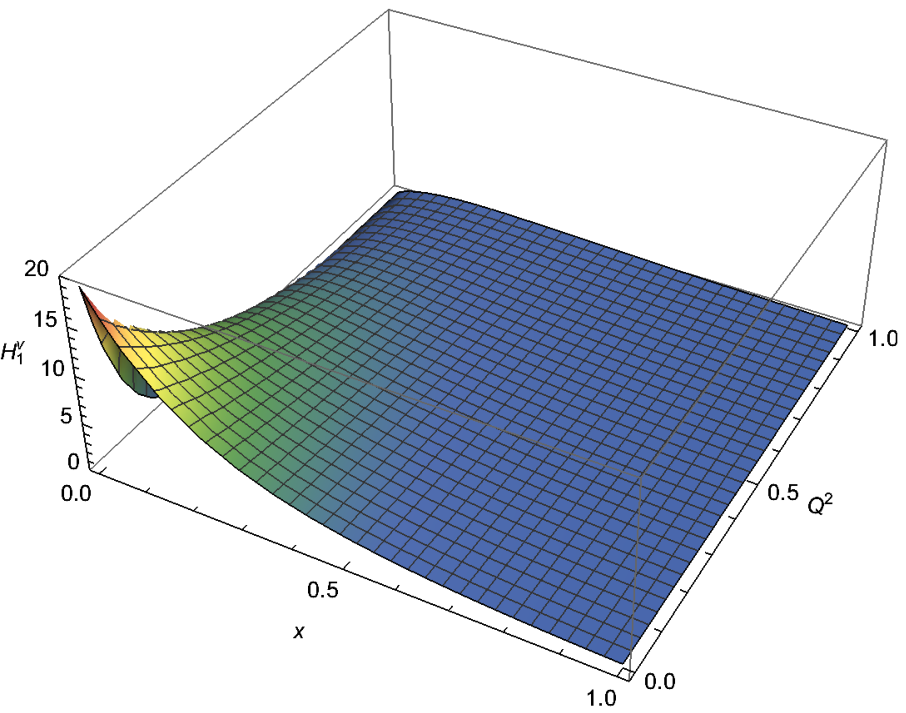}
{(b)}\includegraphics[width=7.5cm,clip]{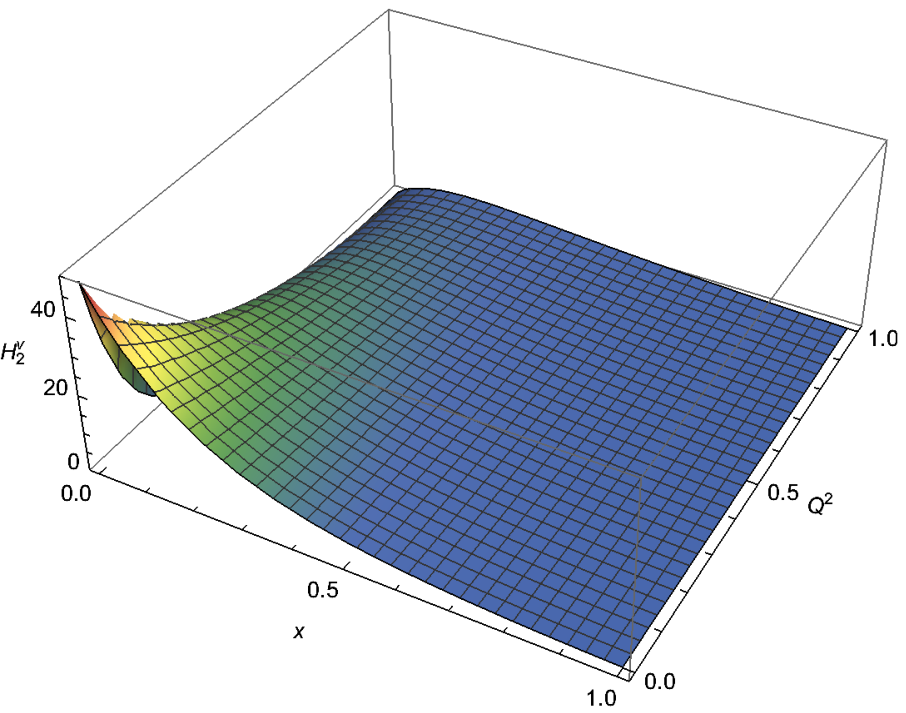}
{(c)}\includegraphics[width=7.5cm,clip]{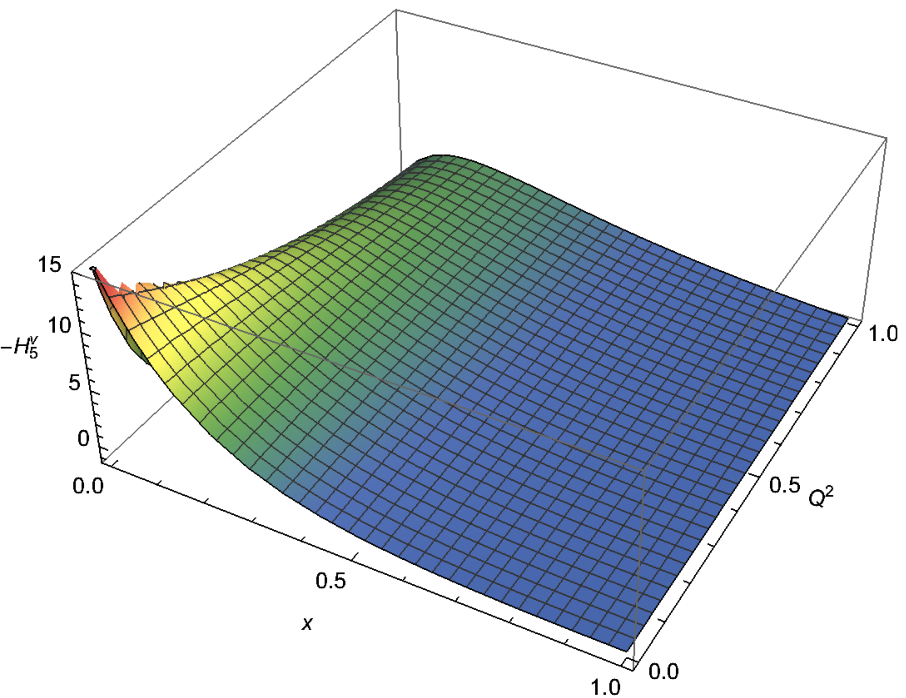}
{(d)}\includegraphics[width=8.5cm,clip]{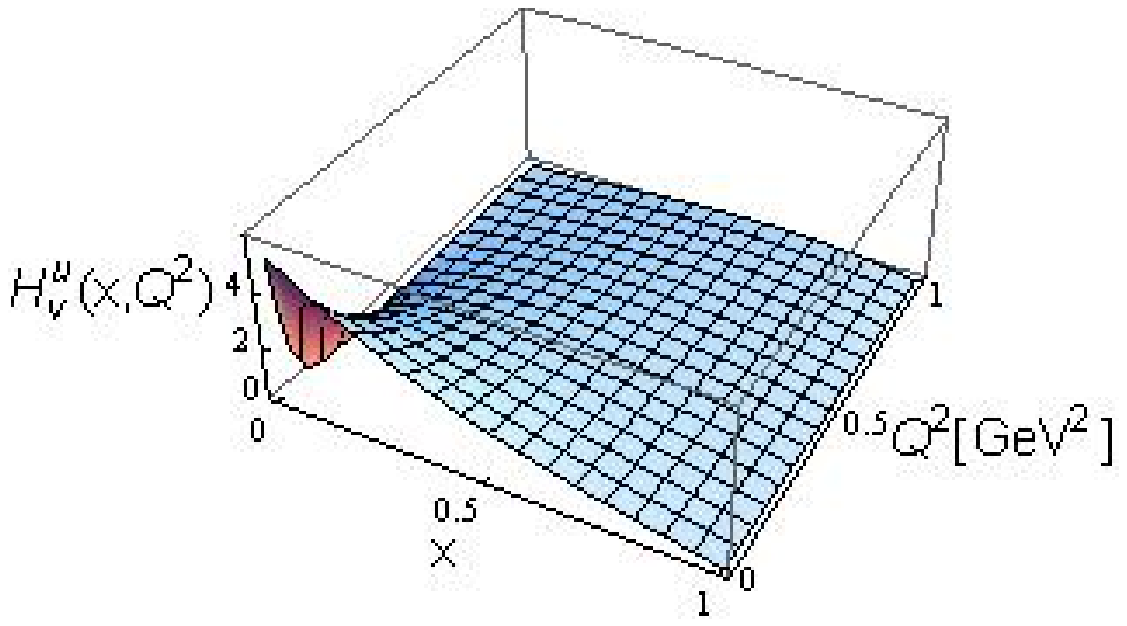}
{(e)}\includegraphics[width=8.5cm,clip]{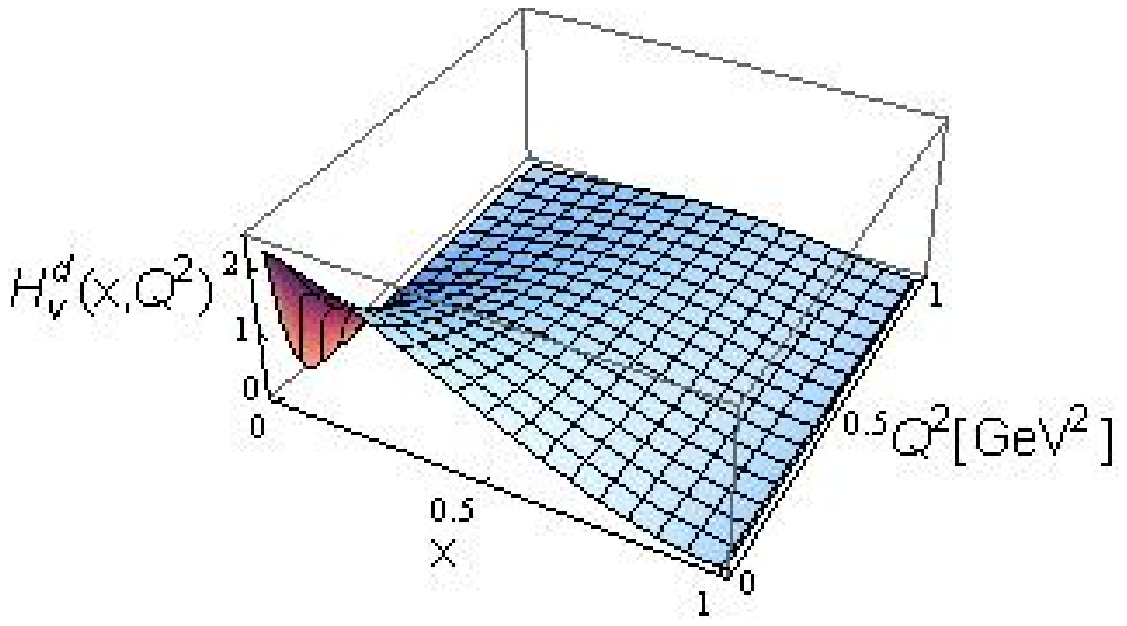}
{(f)}\includegraphics[width=6.5cm,clip]{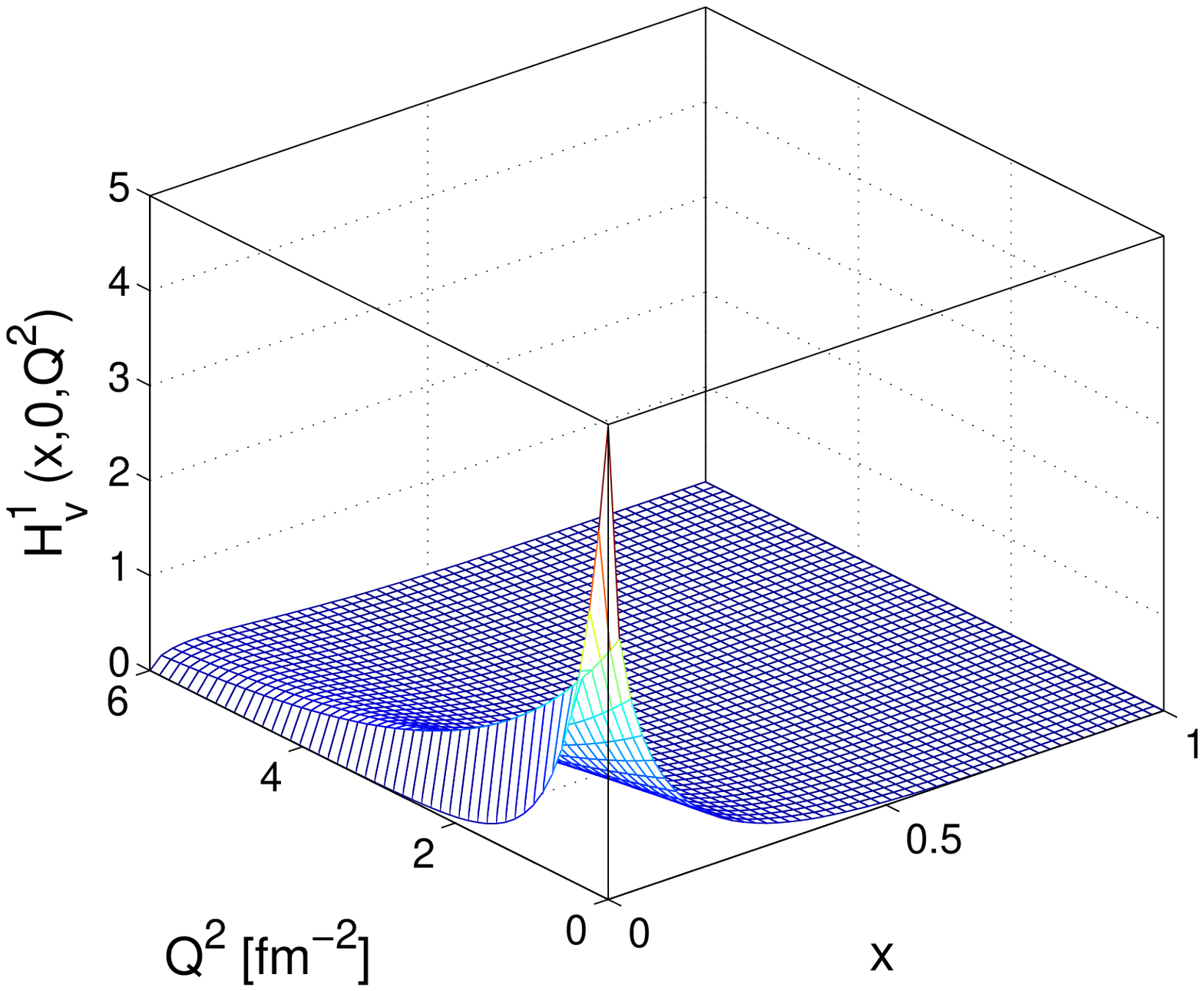}
\end{minipage}
\caption{Plot (a), (b) and (c) are $H_1^v$, $H_2^v$ and $-H_5^v$ GPDs of $\rho$ meson for this work. Plot (d) and (e) are $H_v^u$, $H_v^d$ GPDs for the nucleon of $u$ and $d$ quarks \cite{Vega}. Plot (f) is $H^1_v(x,0,Q^2)$ GPDs for deuteron \cite{Mondal}.}
\label{muqayise}
\end{figure*}

\begin{figure*}[htbp]
\begin{minipage}[c]{0.98\textwidth}
{(a)}\includegraphics[width=6.5cm,clip]{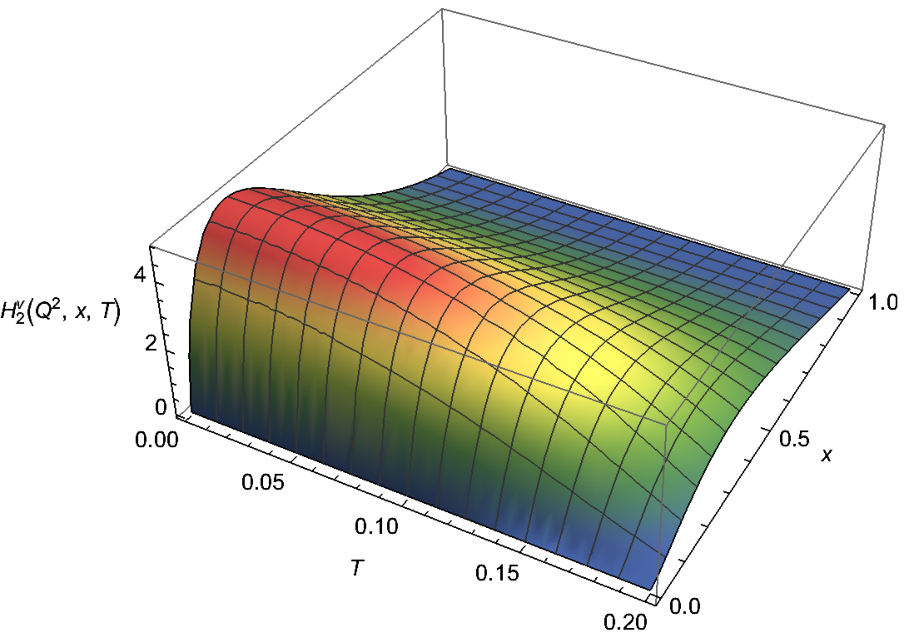}
{(b)}\includegraphics[width=6.5cm,clip]{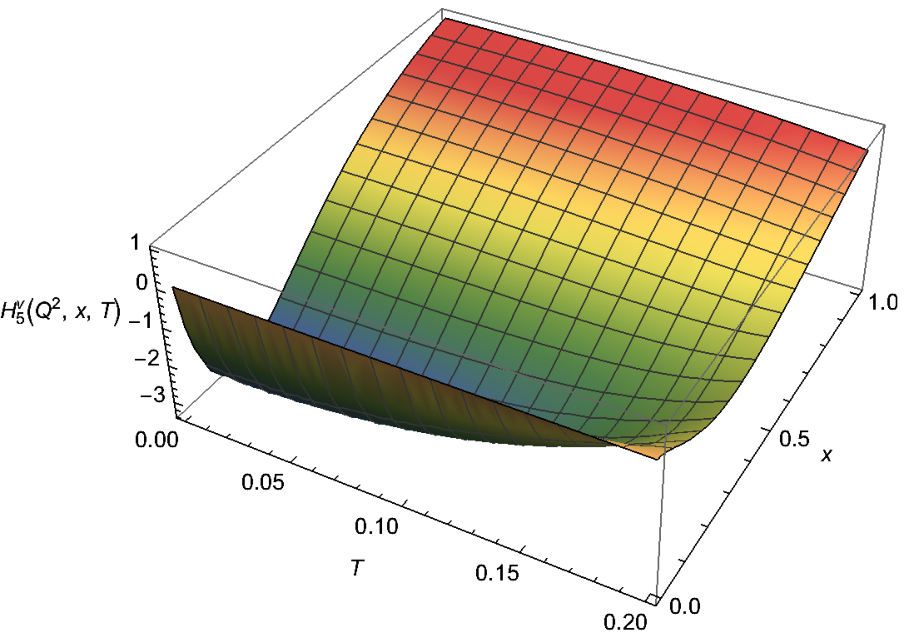}
\end{minipage}
\caption{Plot (a), (b) are generalized parton distributions at finite temperature for $Q^2=1 {\rm\ GeV}$.}
\label{Q_12}
\end{figure*}

\begin{figure*}[htbp]
\begin{minipage}[c]{0.98\textwidth}
{(a)}\includegraphics[width=6.5cm,clip]{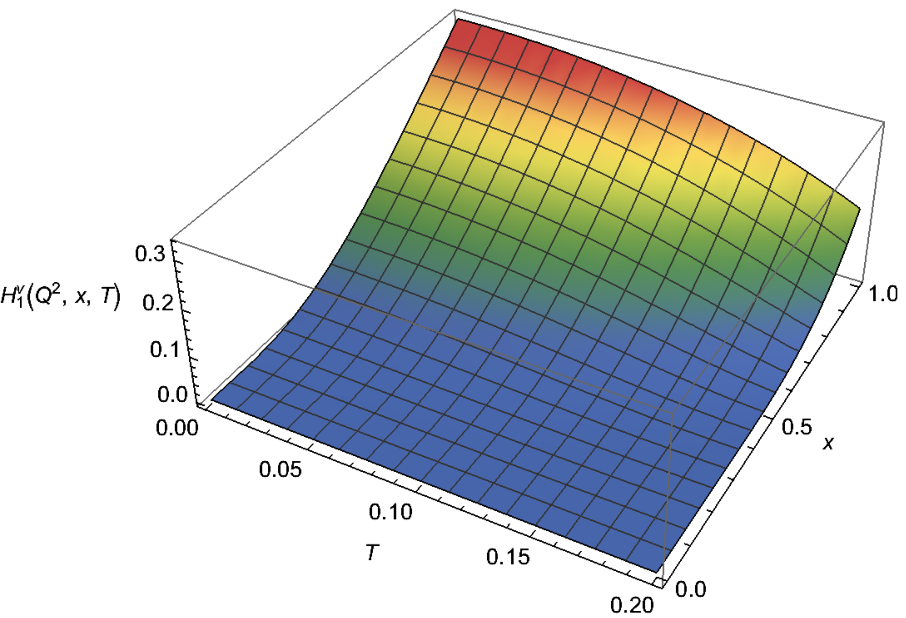}
{(b)}\includegraphics[width=6.5cm,clip]{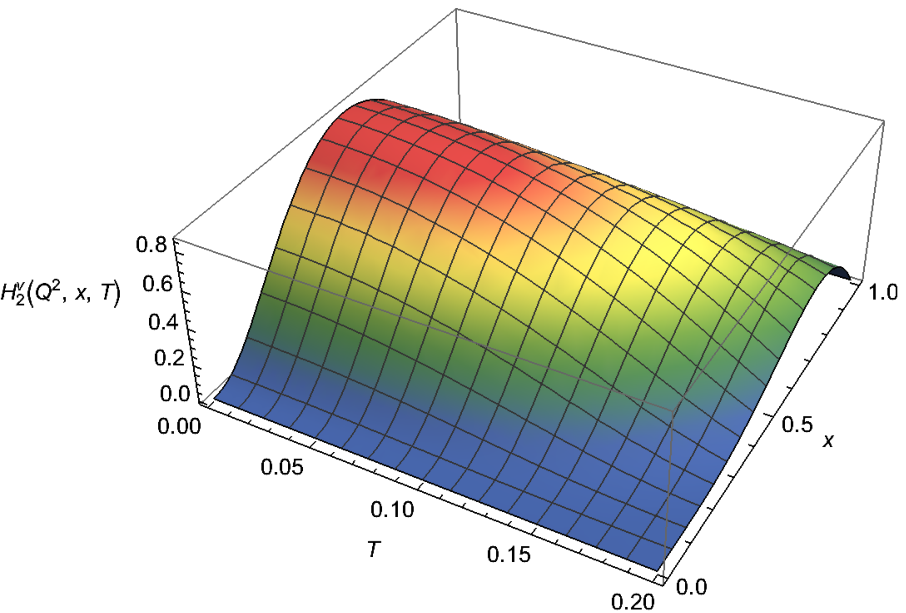}
{(c)}\includegraphics[width=6.5cm,clip]{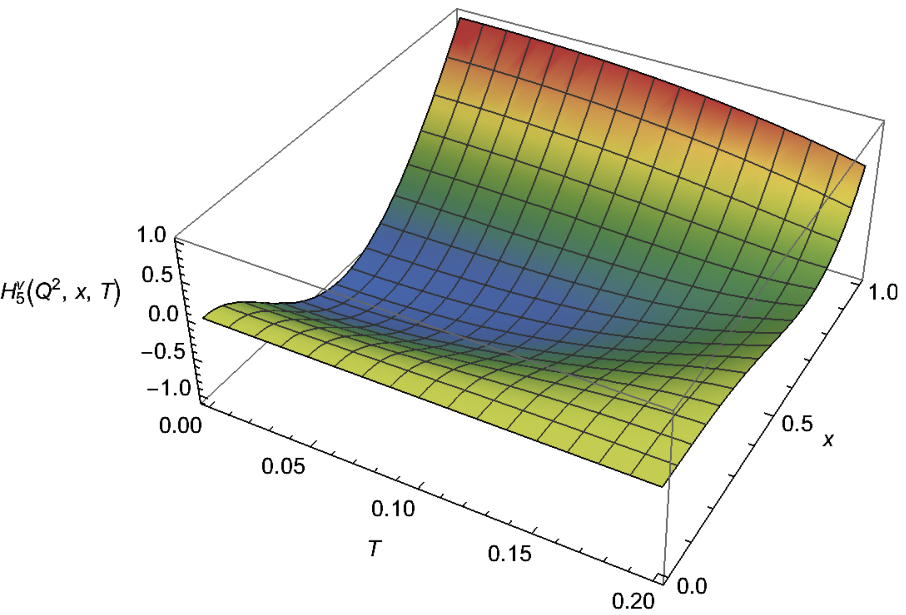}
\end{minipage}
\caption{Plot (a), (b), (c) are generalized parton distributions at finite temperature for $Q^2=3 {\rm\ GeV}$.}
\label{Q_31}
\end{figure*}

\begin{figure*}[htbp]
\begin{minipage}[c]{0.98\textwidth}
{(a)}\includegraphics[width=6.5cm,clip]{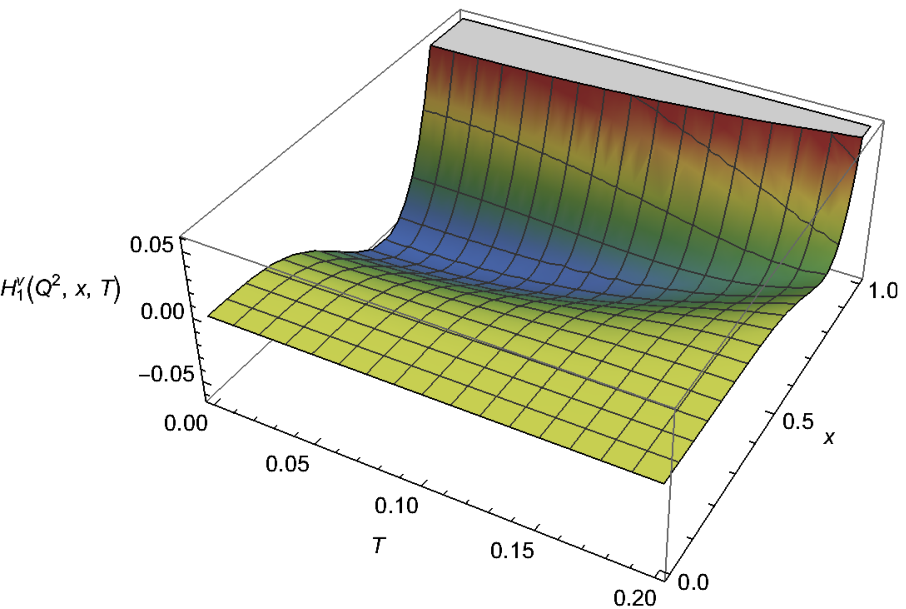}
{(b)}\includegraphics[width=6.5cm,clip]{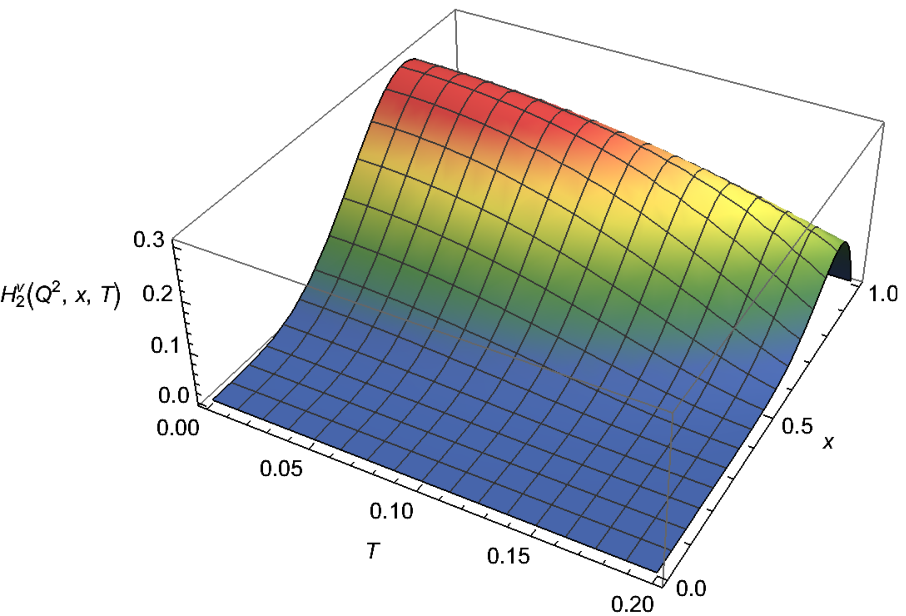}
{(c)}\includegraphics[width=6.5cm,clip]{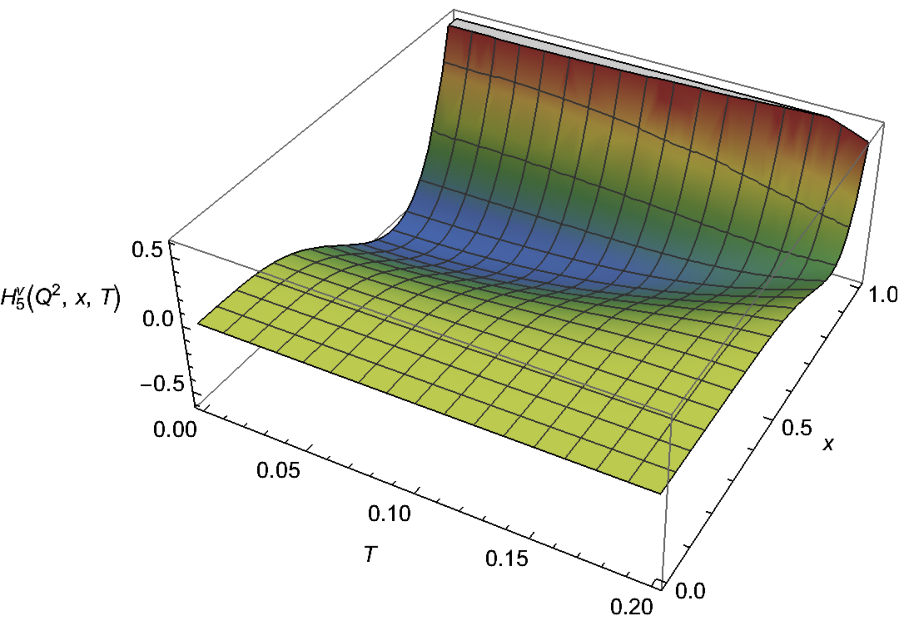}
\end{minipage}
\caption{Plot (a), (b), (c) are generalized parton distributions at finite temperature for $Q^2=6 {\rm\ GeV}$.}
\label{Q_61}
\end{figure*}

In Figs.~\ref{muqayise}-\ref{Q_61} we study the $\rho$ meson’s GPDs at zero and finite temperature for the $\xi=0$ case. In Fig.~\ref{muqayise} (a)-(c) GPDs $H^v_i$ were plotted at zero temperature for the $\rho$ meson. As is seen from these figures the value of GPDs grows when $x \to 0$ and $Q^2 \to 0$. Such behaviour is characteristic of GPDs; similar GPDs graphs' shapes were obtained in Ref.~\cite{VegaGPD} and Ref.~\cite{Mondal}, where $H_i$ were plotted for the nucleon and deuteron, correspondingly. For an apparent comparison, we present in Fig.~\ref{muqayise} (d) and (e)  graphs plotted in \cite{VegaGPD} for the generalised $u$ and $d$ quark distributions $H^{u,d}_v$ in the nucleon and in Fig.~\ref{muqayise} $H^1_v$ obtained in the soft-wall model. This comparison shows the parton distributions on  $x$ in the $\rho$ mesons and nucleons and deuteron are similar. 

GPDs at finite temperature were plotted In Figs.~\ref{Q_12}, \ref{Q_31}, \ref{Q_61} GPDs for the different fixed values $Q^2=1,3,6 {\rm\ GeV}$, correspondingly. As is seen from these plots, the graphs are more sensitive to temperature in area $x \to 0$ than in area $x \to 1$. Also, we observe peaks of GPDs close to $x=1$ point. The value of peaks grows in $Q^2$ growth.

\section{SUMMARY} \label{summary}

Within the soft-wall AdS/ QCD model having thermal dilaton we numerically investigate the temperature dependences of the gravitational form factors and generalized parton distributions of the vector meson. At first, we check an agreement of the zero–temperature GFFs results obtained within the soft- and hard-wall models and find a good agreement. Plots for GPDs show the similarity of distributions of partons inside $\rho$ meson, nucleon and deuteron. Next, within the framework of the thermal dilaton approach in the soft-wall model, we analyze the explicit integral expressions obtained for the temperature dependence and $Q^2$ of the GFFs and GPDs.  The graphs show a decrease in GFFs in temperature increase. Similar decrease we observe in gravitational radius. Finally, the zero-skewness GPD plots were analyzed at zero and finite temperature cases. Some peaks of $H_i$ were observed when $x\to 1$ and these peaks become smaller in value as the temperature grows. It is reasonable to investigate vector mesons' GFFs and GPDs in the top-down approach as well.

\newpage

\begin{figure*}[htbp]
\begin{minipage}[c]{0.98\textwidth}
{(a)}\includegraphics[width=7.5cm,clip]{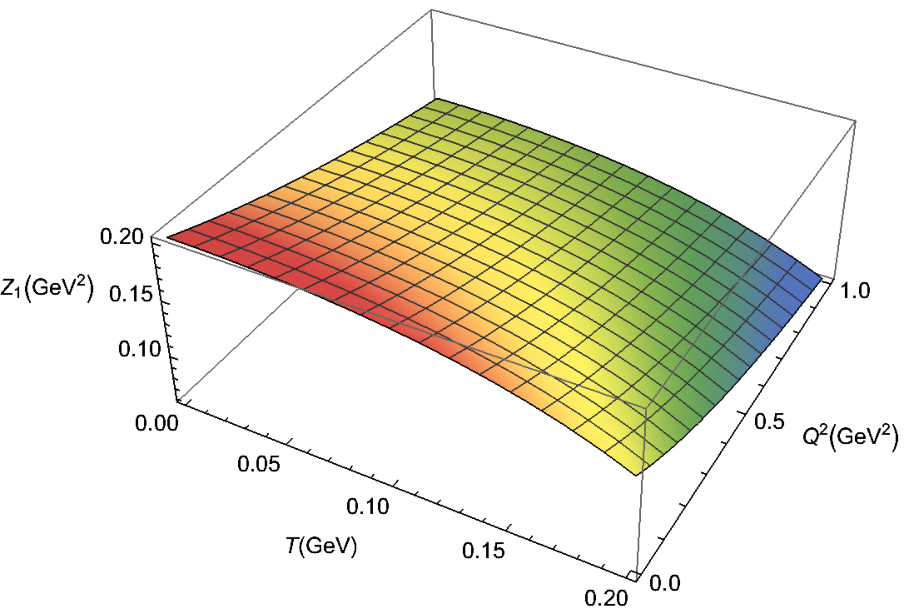}
{(b)}\includegraphics[width=7.5cm,clip]{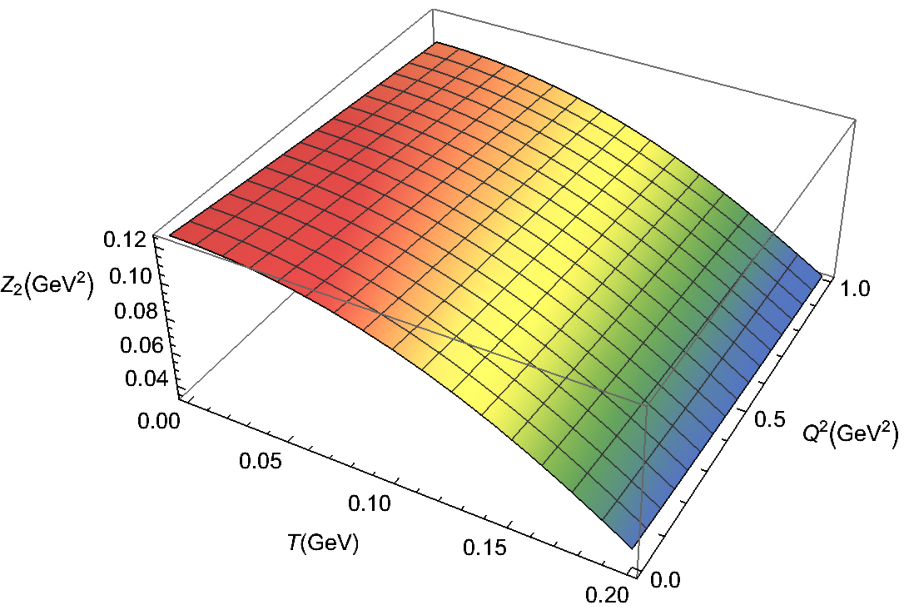}
\end{minipage}
\end{figure*}
\begin{figure}[ht!]
    \caption{Plots of gravitational form factors at finite temperature.}
    \label{Soft_Z1Z2}
\end{figure}

\end{document}